# Single photon wave-front characterization by optical mixing

Shlomo Ruschin*

*Department of Physical Electronics, Faculty of Engineering, Tel-Aviv University, Ramat Aviv Tel-Aviv 69978 Israel.*

**Abstract**

Optical mixing experiments show the ability of amplifying a weak optical signal by superposing it with a stronger one. In the present communication I suggest that the sensitivity of optical mixing can be further enhanced as to allow sensing the wavefront of a single photon's mode simultaneously at two or more locations. Key conditions for that detection is reducing the active size of the detectors below the typical size of the transverse mode, and performing an optical intensity correlation measurement of the Hanbury Brown and Twiss type to the mixed signal. Following these conditions, a macroscopic correlation signal extraction for a single photon emission event is achievable, out of which the photon wave-front characterization at more than one location is attainable with good fidelity. A basic scheme for the demonstration of the effect is proposed and analyzed based on a simple quantum model. An extension of the idea would allow the experimental profiling of an entire wavefront based on a single-photon event, with implications on the issue of the photon wave-particle duality. The validity of the model is confirmed by comparison with previous theoretical and experimental reports involving single photon sources.

Single photons are nowadays a preferred building block for an increasing number of quantum-information related technologies, and their generation, propagation and detection are vigorously pursued [1]. The question of location of a photon in more than one place simultaneously has driven plenty of research wherever the wave-particle nature of photons is manifested. In most reported situations, if a truly single photon source was implemented, the final experimental detection involved also a single photon detection event, and in order to characterize spatially the photon's field distribution at an extended area, the statistical monitoring of multiple events was required [2-4]. Measurements involving a single photon event however are attracting renewed activity and debate nowadays: Piacentini et Al. [5]

demonstrated experimentally a quantum protective measurement involving a single photon, while another report pertain the question of the information content amount that can be carried by a single photon. Tentrup et Al. [6] demonstrated multi-bit transmission via a single photon by angular steering of one photon after its generation. Indeed, the quantum description of a photon field involves also additional classical degrees of freedom, like linear and angular momentum, for which no fundamental quantum restriction exists. The modal content of an optical field also presents a method for imprinting information [7], and in the present communication it is proposed as viable for a single photon event. Extended sensing of a single photon was suggested by entangling an incoming photon with a quantum metamaterial sensor array [8] but no spatial profiling the wavefront was attempted insofar. In the present communication, I suggest transverse optical mode profiling based on a Hanbury-Brown and Twiss (HBT)-type interferometer in which a single photon field is mixed with a strong macroscopic optical field. It is theoretically demonstrated that in such an arrangement, a macroscopic correlation signal can be generated in which the wavefront is manifested at two or more different points. The correlation signal depends on the local amplitude value of the field at the detectors' position, implying that an entire transverse mode can be quantitatively de-convoluted in a single photon event.

Fig.1. depicts the basic proposed scheme. This is apparently associated with photonic correlation arrangements involving a single or multiple photon sources [9]. A prominent previously reported arrangement, closest to the one presented here was demonstrated by Rarity et Al. [10], in which photons from different sources were combined to show a non-

classical anti-correlation Ou-Hong-Mandel (HOM) dip. The fact that quantum interference effects are present for photon from independent sources was predicted back in 1983 by Mandel [11], and further-on demonstrated in many reports [9,12]. Choosing [10] as a reference setup, in which a single-photon field was mixed with an highly attenuated Coherent State (CS) ($|\alpha|^2 < 5$), the setup proposed here differs from it in several aspects: First, the attenuated CS port is replaced by a general highly populated optical state, that will be called as customary Local Oscillator (LO). The second modification is that the transverse spatial mode excited by the single photon is converted from a basic Gaussian into a higher one. This is proposed to emphasize the local character of the detection. The third and main modification is that in the scheme proposed here, the detectors' area is small as compared to the mode cross-sections of the beams involved, sampling a partial section of the mode. The fact that the addition of a single photon to a highly populated CS can drastically modify its properties has been reported many years ago [1213] and is also employed here.

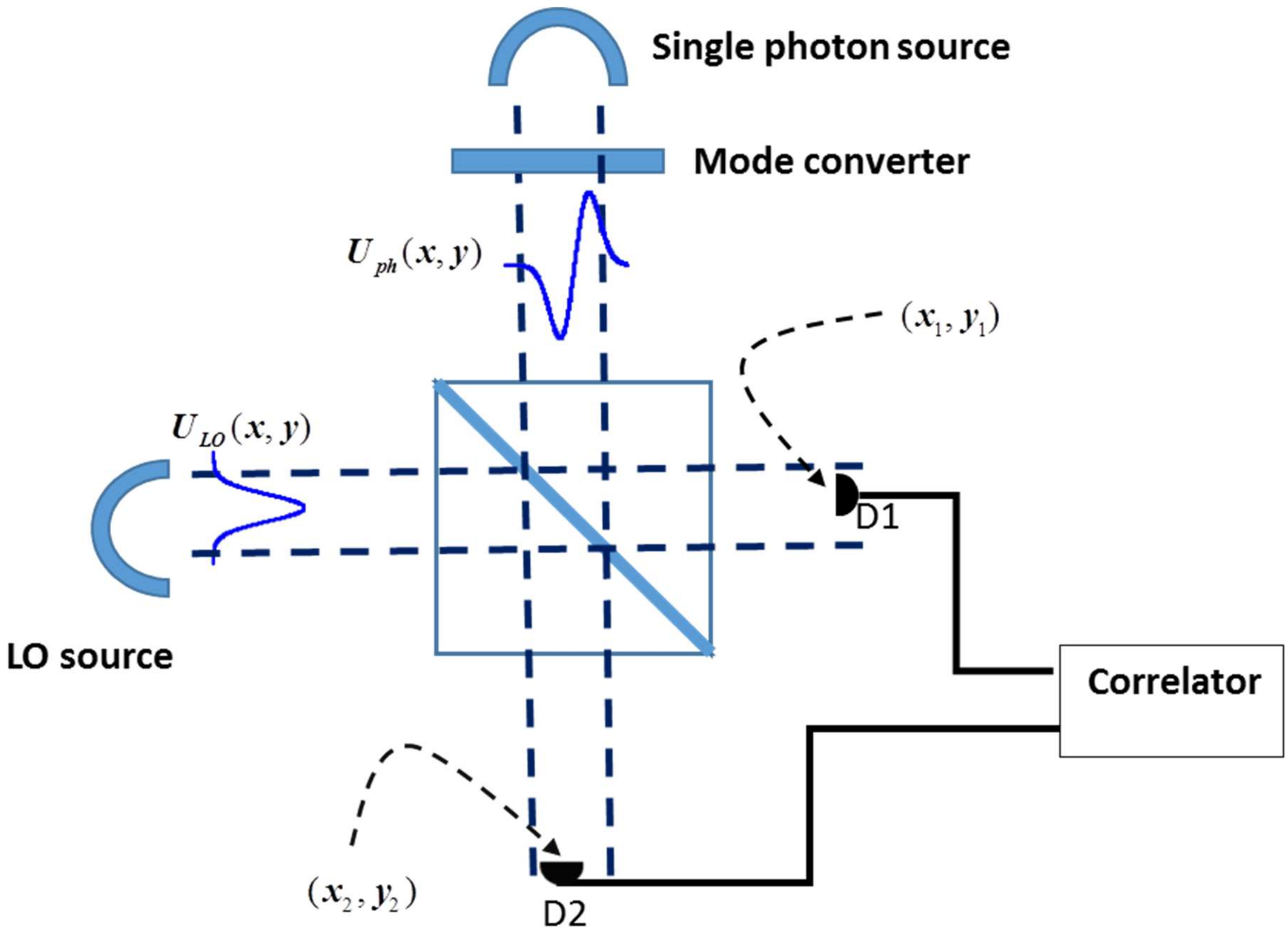

**Figure 1: The proposed basic scheme. Two mutually coherent optical sources with different modal profiles are mixed by a beam-splitter. One source emits a single photon and the second is a higher intensity Local Oscillator (LO) optical field. At the two output ports, detectors are placed exposing small apertures situated at arbitrary transverse locations ($x_1$,$y_1$) and ($x_2$,$y_2$). The detectors are connected to an intensity correlator.**

In the following, the main detectability effect is theoretically demonstrated by a simple quantum model showing explicitly that the intensity correlation signal carries information on the local electric field of the two interacting modes, meaning that if the reference mode is known, the photon mode can be determined. The result is then generalized, by replacing one detector by an array, and performing the correlation measurements simultaneously rendering eventually the entire transverse profile of the photon's mode. For confirmation of the modelling vs, previous theoretical and experimental reports, the formalism is adapted to calculate the correlation signal for the case the detectors are wide but the optical beams are axially displaced.

A standard full quantum approach is adopted for the analysis of the scheme [4, 14, 15]. We start by defining field operators for the positive frequency part of the two electric fields:

$$\hat{E}^{+}_{ph}(\vec{r},t) = A_{ph}(\vec{r}) \cdot \hat{a}_{ph} \qquad \hat{E}^{+}_{LO}(\vec{r},t) = A_{LO}(\vec{r}) \cdot \hat{a}_{LO} \quad , \tag{1}$$

where $\hat{a}_{ph}$ and $\hat{a}_{LO}$ are the usual annihilation operators for the photon and local oscillator fields and $A_{ph}(\vec{r})$, $A_{LO}(\vec{r})$ are vector functions carrying the spatial modal and polarization information of the respective fields, namely:

$$A_{ph}(\vec{r}) = \varepsilon_{ph} U_{ph}(x,y) e^{-ik^{z}_{ph}\cdot z}, \qquad A_{LO}(\vec{r}) = \varepsilon_{LO} U_{LO}(x,y) e^{-ik^{z}_{LO}\cdot z} \tag{2}$$

The modal functions $U_{ph,LO}(x,y)$ express the transverse coordinates' dependencies, which are explicitly different for the photon, and LO fields. $\varepsilon_{ph,LO}$ are the basic electric field unit vectors associated with one photon. A monochromatic propagating mode was assumed for both waves and their time dependence is implicit in the annihilation operators $\hat{a}_{ph}$ and $\hat{a}_{LO}$ (Heisenberg Picture). We further assume the two polarizations to be identical and that the fields share the same angular frequency $\omega$. The two source fields are combined in the beam splitter, represented as a general S-matrix:

$$\begin{bmatrix} \hat{E}^{+}_{1} \\ \hat{E}^{+}_{2} \end{bmatrix} = \begin{bmatrix} s_{11} & s_{12} \\ s_{21} & s_{22} \end{bmatrix} \begin{bmatrix} \hat{E}^{+}_{LO} \\ \hat{E}^{+}_{ph} \end{bmatrix} \tag{3}$$

The fourth-order field correlation rate is next calculated for partial detectors of area $dS$ located in the detectors' planes at positions $(x_1,y_1)$ and $(x_2,y_2)$ respectively:

$$dw^{(2)}(r_1,r_2) = \eta^2 \cdot (dS)^2 \sigma(r_1,r_2) \ , \qquad \sigma(r_1,r_2) = \left\langle \hat{E}^{-}_{1}(r_1)\hat{E}^{-}_{2}(r_2)\hat{E}^{+}_{2}(r_2)\hat{E}^{+}_{1}(r_1) \right\rangle \tag{4}$$

Where $\eta$ is a common efficiency factor, defined for the case of fully open detecting elements, and $dS$ is the active area of each detector. This last expression represents the photocurrent cross-correlation for square-law photodetectors and in a photon-counting scenario it is proportional to the photon coincidence count rate. Considering as base function for the expectation-value calculation $\left|\psi_{LO}\right\rangle\left|1_{ph}\right\rangle$ (uncorrelated input signals), the correlation function $w^{(2)}$ is evaluated by replacing (1) and (2) into (4) (see Supplementary Material for details), furnishing:

$$\sigma(r_1,r_2)=\sigma_{LO}(r_1,r_2)+\sigma_{Mix}(r_1,r_2),$$
$$\sigma_{LO}(r_1,r_2)=\{s_{11}s_{21}A_{LO}(\vec{r}_1)A_{LO}(\vec{r}_2)|^2\left\langle n_{LO}(n_{LO}-1)\right\rangle, \qquad \sigma_{Mix}(r_1,r_2)=|s_{11}s_{22}A_{LO}(\vec{r}_1)A_{ph}(\vec{r}_2)+s_{12}s_{21}A_{LO}(\vec{r}_2)A_{ph}(\vec{r}_1)|^2\left\langle n_{LO}\right\rangle \tag{5}$$

The simplest case of a fully symmetric BS is now replaced namely: $s_{11}=s_{22}=-1/\sqrt{2}$, $s_{12}=s_{21}=i/\sqrt{2}$ rendering:

$$\sigma_{LO}(r_1,r_2)=\tfrac{1}{4}|A_{LO}(\vec{r}_1)A_{LO}(\vec{r}_2)|^2\left\langle n_{LO}(n_{LO}-1)\right\rangle, \qquad \sigma_{Mix}(r_1,r_2)=\tfrac{1}{4}|A_{LO}(\vec{r}_1)A_{ph}(\vec{r}_2)-A_{LO}(\vec{r}_2)A_{ph}(\vec{r}_1)|^2\left\langle n_{LO}\right\rangle \tag{6}$$

where $n_{LO}=\hat{a}^{+}_{LO}\hat{a}_{LO}$. Examining Eq. (6), it can be immediately stated that the mixed term depends on the local values of both modal functions $A_{ph}(\vec{r}_1)$ and $A_{ph}(\vec{r}_2)$ in which the coordinates $\vec{r}_1$ and $\vec{r}_2$ correspond to locations of detectors, arbitrarily separated, each at different arms away from the BS. Note that $\sigma_{Mix}$ is macroscopic, as being amplified by $\left\langle n_{LO}\right\rangle$. The derivation of Eq. (6) (See Supplementary Material), contains no approximations beyond the basic assumptions of the model and is therefore expected to be valid for any arbitrary LO input state $|\psi_{LO}\rangle$. Inspecting some cases critically: If the LO input state contains no photons (vacuum), only the single photon is incident on the BS, and no correlation signal is predicted by Eq. (6) as expected [2,4] . The case where the LO state contains also exactly one photon ($n_{LO}=1$), deserves more attention: According to the HOM effect [11] no correlation signal is expected for single photons incident at different ports even if the photons originated at different sources [12]. This apparently contradicts the outcome of Eq. (6), since the anti-correlation expression contained in the mixing term will in general not cancel. It will cancel however if the modes are identical and the detectors are widely open. Then, Eq. (6) is integrated over the variables $(\vec{r}_1,\vec{r}_2)$ , and the integrals cancel. This is actually the situation encountered in most experiments. Strikingly, the incomplete cancellation of the HOM dip, when encountered, was attributed to non-perfect overlap between the modes [14-16]. In ref [16], the misalignment effect for the $n_{LO}=1$ case, was explicitly calculated and associated with measurements. Non-perfect overlap in time domain was analyzed in [17] rendering similar expressions. This issue is further discussed at the end of this article.

Returning now to analyze the situation depicted in Fig. (1) with partially opened apertures, $\sigma_{LO}$ in Eq. (6) corresponds to the case where only the LO field is present. In the limit $\langle n_{LO}\rangle >> 1$ one gets a classical geometrical beam-splitting ratio for the LO term depending on the product of the field power flux at positions $\vec{r}_1$ and $\vec{r}_2$. One should notice that at that limit, this LO term overvalues the second one as being of the order of $\langle n_{LO}^2\rangle$. This background term cannot be simply subtracted from the signal since its generated noise could made single event detection unrealizable. It is a consequence of the unbalanced disposition of the proposed scheme. The critical issue is quantitatively addressed later on.

For further analysis of the effect of observing a single photon's wavefront in multiple points, we isolate the mixed term of eq. (6) and annotate it in a slightly different form:

$$\sigma_{Mix}(r_1,r_2)=\frac{\varepsilon^4}{4}\,|\,U_{LO}(x_1,y_1)U_{ph}(x_2,y_2)-U_{LO}(x_2,y_2)U_{ph}(x_1,y_1)\,|^2\,\langle n_{LO}\rangle \qquad (7)$$

We note here first that this term has no explicit time dependence as expected after assuming $\omega_{ph}=\omega_{LO}$. Furthermore, we approximated $k^z_{ph}\approx k^z_{LO}$, cancelling also the *z* dependence for travelling wave modes of the same frequency. In practical terms, this means spatial longitudinal beat length [14] is neglected as being long with respect to lengths involved in an experimental situation. The strongest dependence remains on the transversal positions of both detectors $(x_1,y_1)$ and $(x_2,y_2)$. This correlation term will cancel if $x_1=x_2$, $y_1=y_2$ meaning the two points are equivalently located in their respective planes with respect to the modes' profiles (see Fig. 2). It would find its maximal expression if modes $U_{ph}$ and $U_{LO}$ have opposite symmetries. The main point of emphasis is that *both* partial detectors simultaneously contribute to the mixed signal expressed in eq. (7). If we leave one aperture fixed before detector $D_1$ at $(x_1,y_1)$ and scan the second point $(x_2,y_2)$ a graph can be generated as illustrated in Fig.2 from which the profile of the photon's mode can be uniquely deconvolved.

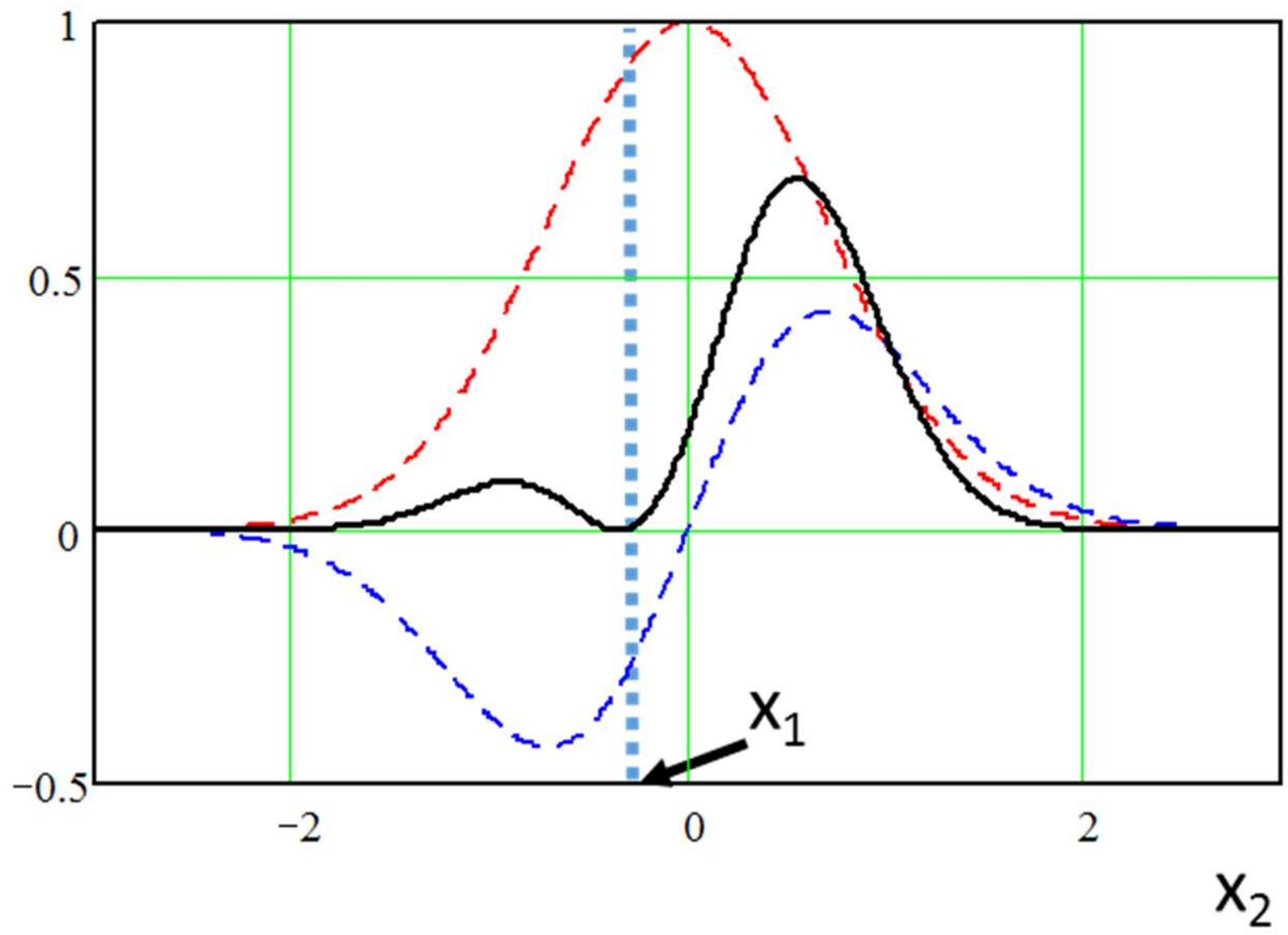

**FIG. 2. The full (black) curve represents the intensity correlation function from apertures located at different detectors with openings at lateral positions $x_1$ (fixed) and $x_2$ (varying). The dashed curves represent the two modes superposed at each detector corresponding to the LO ($TEM_{00}$ ) and Photon ($TEM_{10}$) modes respectively**

Moreover one can envisage a situation as depicted in Fig. 3, in which instead of the scan, an array of partial detectors $\boldsymbol{D_m}$ is distributed in front of the optical beam at the branch location of detector D2, at positions $\{x_m, y_m\}$ and their output is fed individually to the correlator from which multiple correlation functions $\sigma_{0,m} = \left\langle \hat{E}_0^-(r_0)\hat{E}_m^-(r_m)\hat{E}_m^+(r_m)\hat{E}_0^+(r_0)\right\rangle$ are simultaneously determined. Since all these signals are scaled-up into the macroscopic regime by the common factor $\langle n_{LO} \rangle$, one can conclude that the entire mode profile $U_{ph}(x, y)$ of the single photon can be measurably retrieved by means one photon emission event.

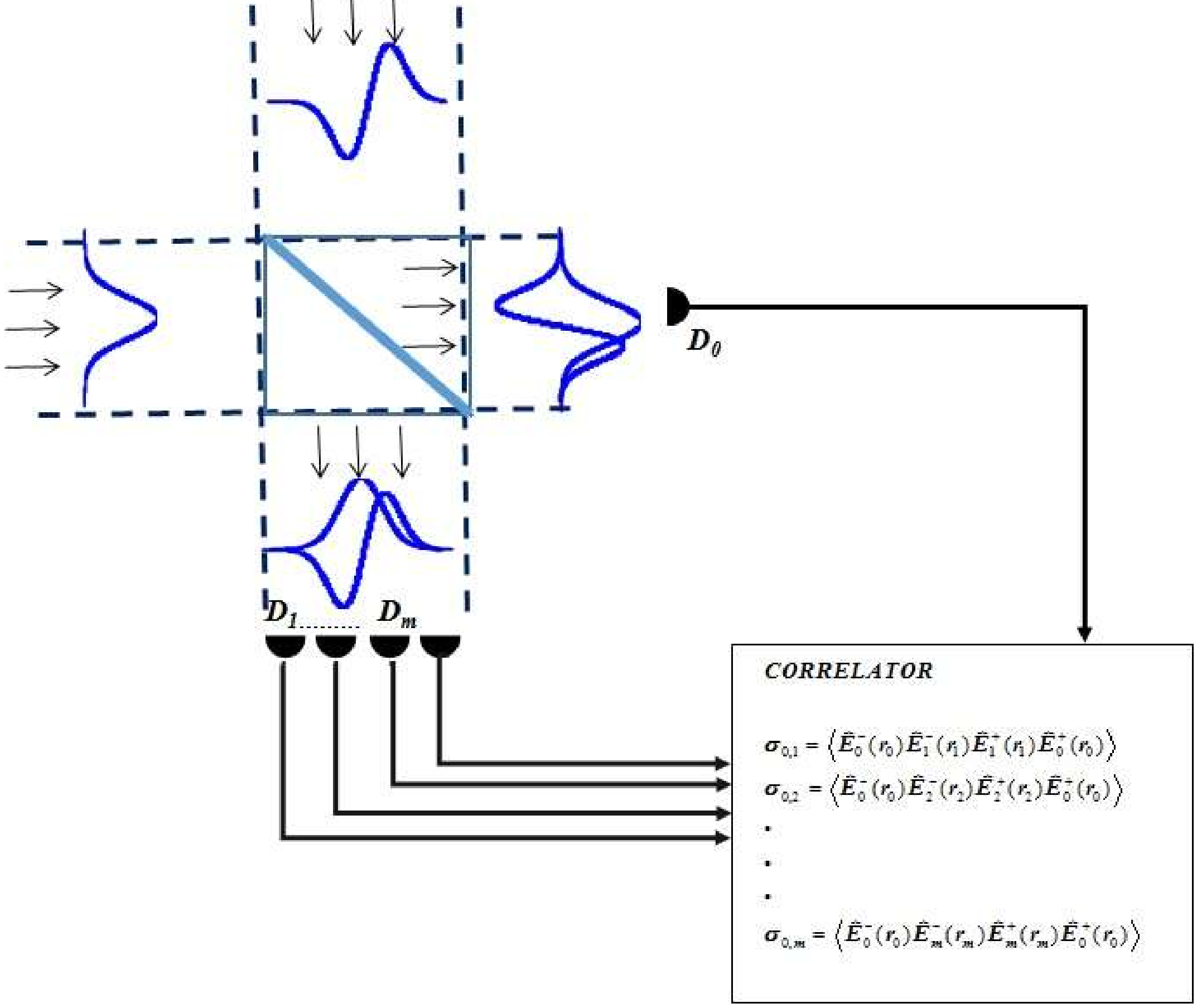

**FIG. 3. Proposed scheme for the simultaneous profiling of the entire photon mode in a single measurement event. The partial detector at one arm of the intensity interferometer is replaced by a detector array. Each detector in the array is independently fed into the correlator**

Regarding the time dependence or spectral content of the electromagnetic fields involved, a monochromatic (infinite) running wave was assumed for simplicity for both waves. The main results however predicted are expected to be valid for photons emitted in finite-time packets as long as they remain mutually coherent during the detection process [16,17]. The same considerations would apply for sources with slightly different frequencies $\omega_{ph} \neq \omega_{LO}$ , or finite bandwidth [17], as long as time coherence is preserved along the relevant measurement time.

An essential point for considering the applicability of the method proposed here is the noise induced by the $\sigma_{LO}$ term of Eqs. 5,6. This dominating term is a consequence of the essentially unbalanced scheme applied. If this noise is larger than the mixed signal $\sigma_{Mix}$, the mode profiling in a single photon event is not feasible and statistical data accumulation is required. For this purpose, a SNR is defined by comparing the average mixed signal to the mean square deviation (MSD) $\Delta\hat{\sigma}$ of the *total* signal of Eq. 6, namely:

$$SNR = {\langle \sigma_{Mix} \rangle}/{\Delta\hat{\sigma}}, \qquad where, \qquad \Delta\sigma = \sqrt{\langle \sigma^2 \rangle - \langle \sigma \rangle^2}, \qquad \sigma = \sigma_{LO} + \sigma_{Mix} \tag{8}$$

This last expression is evaluated explicitly in the Supplementary Material section for the case of a LO in the form of a Coherent State namely an input state: $|\alpha_{LO}\rangle |1_{ph}\rangle$. As the detailed calculation shows, the correlation $\langle \hat{\sigma}_{mix} \rangle$ scales as $|\alpha|^4$ while MSD $\Delta\sigma$ scales as $|\alpha|^3$ leaving an SNR growth proportional to the field strength $|\alpha|$.

As an additional implementation of the formalism developed, the correlation signal for widely opened detectors is calculated by integrating Eq. 6 is over the transverse coordinates. This will allow comparison with established reports on few-photon correlation measurements [10,14,15]. Moreover, the integration allows the introduction of mutual axial displacement [16] between the modal profiles of the photon and local oscillator. In Fig. 4 the effect of misalignment is simulated: the integrated correlation signal is calculated for a shifting profile of the photon's mode while the LO mode is kept fixed at a centered position of the detector window. The width of the detector is ***2d*** and the photon's mode center displacement relative to the detector's center is designated by $\boldsymbol{x_d(ph)}$. Fig 4(a) reproduces the HOM configuration, namely a single photon state is simulated also for LO port, and a zero-dip is attained for full alignment as in refs. [11,15,24], there however the variable was the delay time between the photon pair instead of mutual displacement. The effect non-perfect overlap has been reported to be responsible for the non-complete cancelation of the correlation at the center of the dip [141516]. In Fig. 4(a) also the case of the photon's mode being orthogonal to the one of the LO is shown as a separated trace. This situation has not been considered previously in the literature. When the two orthogonal modes are centered and aligned ($\boldsymbol{x_d(ph)}$=0), the HOM dip is fully filled as expected. At slightly misaligned positions however, ($\boldsymbol{x_d(ph)}\neq 0$), quantum interference effects are still apparent as adjoin dips even for orthogonal modes. Fig. 4(b) displays the case of a LO field being in a populated coherent state. The photon expectation values in Eq. (7) can be then straightforwardly calculated to be $\langle \boldsymbol{n_{LO}} \rangle = |\alpha^2_{LO}|$ and $\langle \boldsymbol{n_{LO}}(\boldsymbol{n_{LO}}-1) \rangle = |\alpha^4_{LO}|$. Furthermore the interference visibility between a coherent state and a single photon was calculated for $0 < |\alpha^2_{LO}| < 5$ and Fig. 2 of ref. [10] was confirmed (see Supplementary Material). Worth remarking is that although the visibility is reduced as a function of the LO strength, the absolute depth of the HOM dip increases.

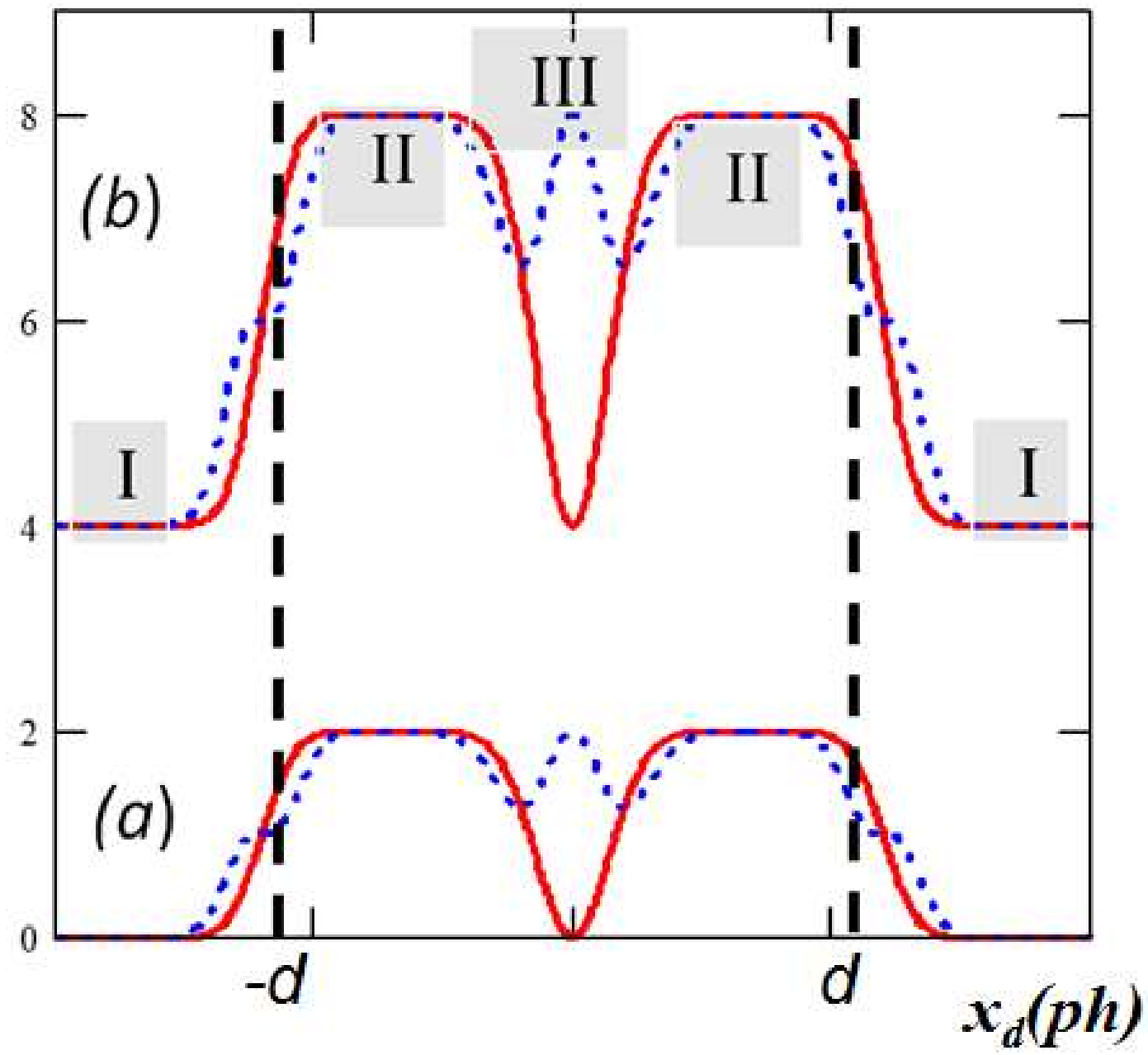

**FIG. 4. Effect of misalignment in fully open detectors: the integrated correlation signal is calculated for a shifting profile of the photon (*Ph)* mode, while the LO mode is kept fixed at a centered position of its detector window. The full (red) lines correspond to interfering $TEM_{00}$ modes for both beams, and the doted (blue) line corresponds to a $TEM_{10}$ mode for the *Ph* mode. In (*a*), a single photon is placed also in the *LO* port, emulating the HOM effect. In (*b*), the single photon is mixed with a coherent state of intensity $|\alpha|^2 = 4$, showing the amplification effect. The vertical lines denote the detectors' aperture. In regions I the *Ph* mode is outside the detector. In II, both modes are inside the detector not overlapping, and in III, partial and full overlap effects are evident.**

Resuming, it is shown that the correlated partial measurement of a single photon event at different spatial locations allows the retrieval of the photon's transverse wavefront. Key enabling features for its observation are exposing the fields involved to small area detectors, and correlating of the photon field with a strong LO field. Second moment calculations for the correlation function demonstrate linear improving of the quantum SNR with the LO field magnitude. An experimental demonstration of the scheme would require employing a highly stable laser as a LO source and close to ideal detectors for reducing non-quantum relative intensity noise (RIN). Alternatively more complex interferometric balanced schemes may be considered [25].

**Supplementary Material for:**

**Single photon wave-front characterization by optical mixing**

## A. Detailed derivation of Eq. (6)

The definition of the entering fields into the BS is:

$$\hat{E}^{+}_{ph}(\vec{r},t) = A_{ph}(\vec{r})\cdot\hat{a}_{ph} \qquad \hat{E}^{+}_{LO}(\vec{r},t) = A_{LO}(\vec{r})\cdot\hat{a}_{LO} \tag{A1}$$

After leaving the BS these fields transform according to Eq. 3 into:

$$\begin{aligned}\hat{E}^{+}_{1}(r_1) &= s_{11}A_{LO}(\vec{r}_1)\cdot\hat{a}_{LO} + s_{12}A_{ph}(\vec{r}_1)\cdot\hat{a}_{ph}\\ \hat{E}^{+}_{2}(r_2) &= s_{21}A_{LO}(\vec{r}_2)\cdot\hat{a}_{LO} + s_{22}A_{ph}(\vec{r}_2)\cdot\hat{a}_{ph}\end{aligned} \tag{A3}$$

To evaluate the correlation function $\left\langle \hat{E}^{-}_{1}(r_1)\hat{E}^{-}_{2}(r_2)\hat{E}^{+}_{2}(r_2)\hat{E}^{+}_{1}(r_1)\right\rangle$, we split the expectation value and calculate first the ket: $\left|\hat{E}^{+}_{2}(r_2)\hat{E}^{+}_{1}(r_1)\right\rangle$, then evaluate its HC and innerly multiply. The evaluation is made in terms of the incoming fields, as the incoming port the photon and LO states are uncoupled and can be expressed as product: $|\psi\rangle = |\psi_{LO}\rangle|\psi_{ph}\rangle$

$$\left|\hat{E}^{+}_{2}(r_2)\hat{E}^{+}_{1}(r_1)\right\rangle = (s_{21}A_{LO}(\vec{r}_2)\cdot\hat{a}_{LO} + s_{22}A_{ph}(\vec{r}_2)\cdot\hat{a}_{ph})(s_{11}A_{LO}(\vec{r}_1)\cdot\hat{a}_{LO} + s_{12}A_{ph}(\vec{r}_1)\cdot\hat{a}_{ph})|\psi\rangle$$

$$\begin{aligned}&= (\underbrace{s_{21}s_{11}A_{LO}(\vec{r}_2)A_{LO}(\vec{r}_1)[\hat{a}_{LO}]^2}_{I} + \underbrace{(s_{22}s_{11}A_{ph}(\vec{r}_2)A_{LO}(\vec{r}_1)\hat{a}_{ph}\hat{a}_{LO} + s_{21}s_{12}A_{LO}(\vec{r}_2)A_{ph}(\vec{r}_1)\hat{a}_{ph}\hat{a}_{LO})}_{II}\\ &\quad \underbrace{+s_{22}s_{12}A_{ph}(\vec{r}_2)A_{ph}(\vec{r}_1)[\hat{a}_{ph}]^2}_{III})|\psi_{LO}\rangle|\psi_{ph}\rangle\end{aligned} \tag{A4}$$

I used here the fact that $\hat{\boldsymbol{a}}_{ph}, \hat{\boldsymbol{a}}_{LO}$ commute.

Now we evaluate separately each of the terms ***I,II,III*** in (A4), making use of the fact that $\left|\psi_{ph}\right\rangle = \left|1_{ph}\right\rangle$ stands for a single photon field, thus $[\hat{\boldsymbol{a}}_{ph}]\left|\psi_{ph}\right\rangle = \left|0_{ph}\right\rangle$ and $[\hat{\boldsymbol{a}}_{ph}]^2\left|\psi_{ph}\right\rangle = 0$
One is left then with:

$$
\begin{aligned}
&\boldsymbol{I} = \boldsymbol{s}_{21}\boldsymbol{s}_{11}\boldsymbol{A}_{LO}(\vec{\boldsymbol{r}}_2)\boldsymbol{A}_{LO}(\vec{\boldsymbol{r}}_1)[\hat{\boldsymbol{a}}_{LO}]^2\left|\psi_{LO}\right\rangle\left|1_{ph}\right\rangle \\
&\boldsymbol{II} = (\boldsymbol{s}_{22}\boldsymbol{s}_{11}\boldsymbol{A}_{ph}(\vec{\boldsymbol{r}}_2)\boldsymbol{A}_{LO}(\vec{\boldsymbol{r}}_1) + \boldsymbol{s}_{21}\boldsymbol{s}_{12}\boldsymbol{A}_{LO}(\vec{\boldsymbol{r}}_2)\boldsymbol{A}_{ph}(\vec{\boldsymbol{r}}_1))[\hat{\boldsymbol{a}}_{LO}\left|\psi_{LO}\right\rangle]\left|0_{ph}\right\rangle \\
&\boldsymbol{III} = 0
\end{aligned}
\tag{A5}
$$

Calculating similarly, $\left\langle \hat{\boldsymbol{E}}_2^-(\boldsymbol{r}_2)\hat{\boldsymbol{E}}_1^-(\boldsymbol{r}_1)\right|$ by Hermitian conjugation and multiplying, considering the orthogonality of the photon functions $\left\langle 1_{ph}\right|\left|0_{ph}\right\rangle = 0$ we are left with only 2 terms:

$$
\begin{aligned}
\left\langle \hat{\boldsymbol{E}}_2^-(\boldsymbol{r}_2)\hat{\boldsymbol{E}}_1^-(\boldsymbol{r}_1)\right|\left|\hat{\boldsymbol{E}}_2^+(\boldsymbol{r}_2)\hat{\boldsymbol{E}}_1^+(\boldsymbol{r}_1)\right\rangle &= |\boldsymbol{s}_{21}\boldsymbol{s}_{11}\boldsymbol{A}_{LO}(\vec{\boldsymbol{r}}_2)\boldsymbol{A}_{LO}(\vec{\boldsymbol{r}}_1)|^2 \left\langle\psi_{LO}\right|[\hat{\boldsymbol{a}}_{LO}{}^\dagger]^2[\hat{\boldsymbol{a}}_{LO}]^2\left|\psi_{LO}\right\rangle \\
&+ |\boldsymbol{s}_{22}\boldsymbol{s}_{11}\boldsymbol{A}_{ph}(\vec{\boldsymbol{r}}_2)\boldsymbol{A}_{LO}(\vec{\boldsymbol{r}}_1) + \boldsymbol{s}_{21}\boldsymbol{s}_{12}\boldsymbol{A}_{LO}(\vec{\boldsymbol{r}}_2)\boldsymbol{A}_{ph}(\vec{\boldsymbol{r}}_1)|^2 \,[\left\langle\psi_{LO}\right|[\hat{\boldsymbol{a}}_{LO}{}^\dagger\hat{\boldsymbol{a}}_{LO}]\left|\psi_{LO}\right\rangle]
\end{aligned}
\tag{A6}
$$

The operators appearing here can be expressed in terms of the number operators of the *LO* field, namely: $\hat{\boldsymbol{a}}_{LO}{}^\dagger\hat{\boldsymbol{a}}_{LO} = \hat{\boldsymbol{n}}_{LO}$ and. After replacing the value of a fully symmetric S matrix: $\boldsymbol{s}_{11} = \boldsymbol{s}_{22} = -1/\sqrt{2}, \quad \boldsymbol{s}_{12} = \boldsymbol{s}_{21} = \boldsymbol{i}/\sqrt{2}$, Eq. 6 of the paper is retrieved:

$$
\left\langle \hat{\boldsymbol{E}}_2^-(\boldsymbol{r}_2)\hat{\boldsymbol{E}}_1^-(\boldsymbol{r}_1)\right|\left|\hat{\boldsymbol{E}}_2^+(\boldsymbol{r}_2)\hat{\boldsymbol{E}}_1^+(\boldsymbol{r}_1)\right\rangle = \tfrac{1}{4}\{|\boldsymbol{A}_{LO}(\vec{\boldsymbol{r}}_1)\boldsymbol{A}_{LO}(\vec{\boldsymbol{r}}_2)|^2\left\langle\hat{\boldsymbol{n}}_{LO}(\hat{\boldsymbol{n}}_{LO}-1)\right\rangle + |\boldsymbol{A}_{LO}(\vec{\boldsymbol{r}}_1)\boldsymbol{A}_{ph}(\vec{\boldsymbol{r}}_2) - \boldsymbol{A}_{LO}(\vec{\boldsymbol{r}}_2)\boldsymbol{A}_{ph}(\vec{\boldsymbol{r}}_1)|^2\left\langle\hat{\boldsymbol{n}}_{LO}\right\rangle\}
\tag{A7}
$$

## B. Calculation of the moments of correlation functions by normal ordering

The aim here is to calculate the variance:

$$
\Delta\hat{\sigma} = \sqrt{\left\langle\hat{\sigma}^2\right\rangle - \left\langle\hat{\sigma}\right\rangle^2}
$$

Meaning the calculation of the first and second moment of the correlation function $\sigma$. Start by decomposing it into a product (cf. Eq. 4)

$\hat{\sigma} = \hat{\sigma}_{12}^-\hat{\sigma}_{12}^+$ where:

$$
\hat{\sigma}_{12}^+ = \hat{\boldsymbol{E}}_2^+(\boldsymbol{r}_2)\hat{\boldsymbol{E}}_1^+(\boldsymbol{r}_1) = \tfrac{1}{2}(\boldsymbol{i}\boldsymbol{A}_{LO}(\vec{\boldsymbol{r}}_2)\cdot\hat{\boldsymbol{a}}_{LO} - \boldsymbol{A}_{ph}(\vec{\boldsymbol{r}}_2)\cdot\hat{\boldsymbol{a}}_{ph})(-\boldsymbol{A}_{LO}(\vec{\boldsymbol{r}}_1)\cdot\hat{\boldsymbol{a}}_{LO} + \boldsymbol{i}\boldsymbol{A}_{ph}(\vec{\boldsymbol{r}}_1)\cdot\hat{\boldsymbol{a}}_{ph})
\tag{B1}
$$

In the following, a coherent state is assumed for the LO field at the incoming port.
$\left(\left|\psi\right\rangle = \left|\alpha_{LO}\right\rangle \left|1_{ph}\right\rangle\right)$

For the first moment $\left\langle\widehat{\sigma}\right\rangle$, a CS replaces the LO in Eq. (A7), rendering immediately:

$$\left\langle\widehat{\sigma}\right\rangle = \tfrac{1}{4}\left|(A_{LO}(\vec{r}_1)A_{ph}(\vec{r}_2) - A_{ph}(\vec{r}_1)A_{LO}(\vec{r}_2))\right|^2 \left|\alpha_{LO}\right|^2 + \tfrac{1}{4}\left|A_{LO}(\vec{r}_2)A_{LO}(\vec{r}_1)\right|^2 \left|\alpha_{LO}\right|^4 \tag{B4}$$

The second order moment $\left\langle\widehat{\sigma}^2\right\rangle$, poses a more extended calculation challenge, being an eight-order correlation function of the detected fields $\left\{\widehat{\boldsymbol{E}}^{\pm}_{1,2}(\boldsymbol{r}_{1,2})\right\}$. These fields are expressed as sums of the emitted fields $\widehat{\boldsymbol{E}}^{\pm}_{\boldsymbol{LO}}(\vec{\boldsymbol{r}},\boldsymbol{t}), \widehat{\boldsymbol{E}}^{\pm}_{\boldsymbol{Ph}}(\vec{\boldsymbol{r}},\boldsymbol{t})$ in whose terms 64 products need to be evaluated. Adopting a Normal Ordering (NO) strategy, simplifies to a considerable extent the task.

Decomposing the second order moment: $(\widehat{\sigma}^{-}_{12}\widehat{\sigma}^{+}_{12})^2 = (\widehat{\sigma}^{-}_{12}\widehat{\sigma}^{+}_{12}\widehat{\sigma}^{-}_{12}\widehat{\sigma}^{+}_{12})$, one notices that as defined it is not normally ordered.

First, I define the commutators:

$$\left[\widehat{\boldsymbol{E}}^{+}_{i}(\boldsymbol{r}_i), \widehat{\boldsymbol{E}}^{-}_{j}(\boldsymbol{r}_j)\right] = \boldsymbol{B}_{ij}, \qquad \boldsymbol{i},\boldsymbol{j} = 1,2 \tag{B5}$$

None of these commutators is zero, as they denote fields after the beam splitter and entangle the LO and *Ph* input states. Their values depend exclusively on the local values of the modal shapes at the detectors' positions, according to A3 :

$$\begin{aligned} &\boldsymbol{B}_{11} = (1/2)(\left|\boldsymbol{A}_{\boldsymbol{LO}}(\vec{\boldsymbol{r}}_1)\right|^2 + \left|\boldsymbol{A}_{\boldsymbol{ph}}(\vec{\boldsymbol{r}}_1)\right|^2), \qquad &&\boldsymbol{B}_{22} = (1/2)(\left|\boldsymbol{A}_{\boldsymbol{LO}}(\vec{\boldsymbol{r}}_2)\right|^2 + \left|\boldsymbol{A}_{\boldsymbol{ph}}(\vec{\boldsymbol{r}}_2)\right|^2) \\ &\boldsymbol{B}_{12} = (1/2\boldsymbol{i})(\boldsymbol{A}_{\boldsymbol{LO}}(\vec{\boldsymbol{r}}_1)\boldsymbol{A}_{\boldsymbol{LO}}(\vec{\boldsymbol{r}}_2)^* - \boldsymbol{A}_{\boldsymbol{ph}}(\vec{\boldsymbol{r}}_1)\boldsymbol{A}_{\boldsymbol{ph}}(\vec{\boldsymbol{r}}_2)^*) \qquad &&\boldsymbol{B}_{21} = \boldsymbol{B}_{12}^* \end{aligned} \tag{B6}$$

The NO expression for the second moment expressed on these terms, furnishes:

$$\begin{aligned} &\left[(\widehat{\sigma}^{-}_{12}\widehat{\sigma}^{+}_{12}\widehat{\sigma}^{-}_{12}\widehat{\sigma}^{+}_{12})\right]_{NO} = \left[\widehat{E}^{-}_2(r_2)\widehat{E}^{-}_1(r_1)\widehat{E}^{+}_2(r_2)\widehat{E}^{+}_1(r_1)\widehat{E}^{-}_2(r_2)\widehat{E}^{-}_1(r_1)\widehat{E}^{+}_2(r_2)\widehat{E}^{+}_1(r_1)\right]_{NO} = \\ &\left[(\widehat{E}^{-}_2(r_2)\widehat{E}^{-}_1(r_1))^2(\widehat{E}^{+}_2(r_2)\widehat{E}^{+}_1(r_1))^2\right] + B_{11}\left[\widehat{E}^{-}_1(r_1)\widehat{E}^{-}_2(r_2)\widehat{E}^{-}_2(r_2)\widehat{E}^{+}_2(r_2)\widehat{E}^{+}_2(r_2)\widehat{E}^{+}_1(r_1)\right] \\ &+B_{22}\left[\widehat{E}^{-}_1(r_1)\widehat{E}^{-}_1(r_1)\widehat{E}^{-}_2(r_2)\widehat{E}^{+}_2(r_2)\widehat{E}^{+}_1(r_1)\widehat{E}^{+}_1(r_1)\right] + B_{12}\left[\widehat{E}^{-}_1(r_1)\widehat{E}^{-}_1(r_1)\widehat{E}^{-}_2(r_2)\widehat{E}^{+}_2(r_2)\widehat{E}^{+}_2(r_2)\widehat{E}^{+}_1(r_1)\right] \\ &B_{21}\left[\widehat{E}^{-}_1(r_1)\widehat{E}^{-}_2(r_2)\widehat{E}^{-}_2(r_2)\widehat{E}^{+}_2(r_2)\widehat{E}^{+}_1(r_1)\widehat{E}^{+}_1(r_1)\right] + (B_{11}B_{22} + B_{11}B_{21})\left[\widehat{E}^{-}_1(r_1)\widehat{E}^{-}_1(r_1)\widehat{E}^{+}_2(r_2)\widehat{E}^{+}_1(r_1)\right] \end{aligned} \tag{B7}$$

These are general operator expressions and straightforward, normally ordered facilitating calculation by applying products of $\widehat{E}^{+}_{1,2}(r_{1,2})$ operators to the right ket and products of $\widehat{E}^{-}_{1,2}(r_{1,2})$ operators to the left bra. Furthermore, since the expectation value is evaluated for the state $\left|\alpha_{LO}\right\rangle\left|1_{ph}\right\rangle$, the application of the $\widehat{E}^{+}_{1,2}(r_{1,2})$ any number of times mixes only states within the sub-space spanned by $\left\{\left|\alpha_{LO}\right\rangle\left|1_{ph}\right\rangle,\left|\alpha_{LO}\right\rangle\left|0_{ph}\right\rangle\right\}$. For this specific state, the moments needed for the calculation of $\left\langle 1_{ph}\right|\left\langle \alpha_{LO}\right|(\widehat{\sigma}^{-}_{12}\widehat{\sigma}^{+}_{12})^2\left|\alpha_{LO}\right\rangle\left|1_{ph}\right\rangle$ according to (B7) are listed here:

$$
\begin{aligned}
\langle\left(\hat{E}_2^-(\vec{r}_2)\hat{E}_1^-(\vec{r}_1)\right)^2\left(\hat{E}_2^+(\vec{r}_2)\hat{E}_1^+(\vec{r}_1)\right)^2\rangle &= 1/16\,|A_{\mathrm{LO}}(\vec{r}_1)|^4|A_{\mathrm{LO}}(\vec{r}_2)|^4|\alpha|^8 \\
&\quad +1/4\,|A_{\mathrm{LO}}(\vec{r}_1)A_{\mathrm{ph}}(\vec{r}_2)-A_{\mathrm{ph}}(\vec{r}_1)A_{\mathrm{LO}}(\vec{r}_2)|^2|A_{\mathrm{LO}}(\vec{r}_2)|^2|\alpha|^6, \\
\langle\hat{E}_1^-(\vec{r}_1)\hat{E}_2^-(\vec{r}_2)\hat{E}_2^-(\vec{r}_2)\hat{E}_2^+(\vec{r}_2)\hat{E}_2^+(\vec{r}_2)\hat{E}_1^+(\vec{r}_1)\rangle &= 1/8\,|A_{\mathrm{LO}}(\vec{r}_1)|^2|A_{\mathrm{LO}}(\vec{r}_2)|^4|\alpha|^6 \\
&\quad +1/8\,|2A_{\mathrm{LO}}(\vec{r}_1)A_{\mathrm{ph}}(\vec{r}_2)-A_{\mathrm{ph}}(\vec{r}_1)A_{\mathrm{LO}}(\vec{r}_2)|^2|A_{\mathrm{LO}}(\vec{r}_2)|^2|\alpha|^4, \\
\langle\hat{E}_1^-(\vec{r}_1)\hat{E}_1^-(\vec{r}_1)\hat{E}_2^-(\vec{r}_2)\hat{E}_2^+(\vec{r}_2)\hat{E}_1^+(\vec{r}_1)\hat{E}_1^+(\vec{r}_1)\rangle &= 1/8\,|A_{\mathrm{LO}}(\vec{r}_1)|^4|A_{\mathrm{LO}}(\vec{r}_2)|^2|\alpha|^6 \\
&\quad +1/8\,|A_{\mathrm{LO}}(\vec{r}_1)A_{\mathrm{ph}}(\vec{r}_2)-2A_{\mathrm{ph}}(\vec{r}_1)A_{\mathrm{LO}}(\vec{r}_2)|^2|A_{\mathrm{LO}}(\vec{r}_1)|^2|\alpha|^4, \\
\langle\hat{E}_1^-(\vec{r}_1)\hat{E}_1^-(\vec{r}_1)\hat{E}_2^-(\vec{r}_2)\hat{E}_2^+(\vec{r}_2)\hat{E}_2^+(\vec{r}_2)\hat{E}_1^+(\vec{r}_1)\rangle &= i/8\,\overline{A_{\mathrm{LO}}(\vec{r}_1)}A_{\mathrm{LO}}(\vec{r}_2)(|A_{\mathrm{LO}}(\vec{r}_1)|^2|A_{\mathrm{LO}}(\vec{r}_2)|^2|\alpha|^6 \\
&\quad +2\left|A_{\mathrm{LO}}(\vec{r}_1)A_{\mathrm{ph}}(\vec{r}_2)\right|^2|\alpha|^4+2\left|A_{\mathrm{LO}}(\vec{r}_2)A_{\mathrm{ph}}(\vec{r}_1)\right|^2|\alpha|^4 \\
&\quad -4A_{\mathrm{LO}}(\vec{r}_1)A_{\mathrm{ph}}(\vec{r}_2)\overline{A_{\mathrm{LO}}(\vec{r}_2)}\,\overline{A_{\mathrm{ph}}(\vec{r}_1)}|\alpha|^4 \\
&\quad -\overline{A_{\mathrm{LO}}(\vec{r}_1)}\,\overline{A_{\mathrm{ph}}(\vec{r}_2)}A_{\mathrm{LO}}(\vec{r}_2)A_{\mathrm{ph}}(\vec{r}_1)|\alpha|^4), \\
\langle\hat{E}_1^-(\vec{r}_1)\hat{E}_2^-(\vec{r}_2)\hat{E}_2^-(\vec{r}_2)\hat{E}_2^+(\vec{r}_2)\hat{E}_1^+(\vec{r}_1)\hat{E}_1^+(\vec{r}_1)\rangle &= -i/8\,A_{\mathrm{LO}}(\vec{r}_1)\overline{A_{\mathrm{LO}}(\vec{r}_2)}(|A_{\mathrm{LO}}(\vec{r}_1)|^2|A_{\mathrm{LO}}(\vec{r}_2)|^2|\alpha|^6 \\
&\quad +2\left|A_{\mathrm{LO}}(\vec{r}_1)A_{\mathrm{ph}}(\vec{r}_2)\right|^2|\alpha|^4+2\left|A_{\mathrm{LO}}(\vec{r}_2)A_{\mathrm{ph}}(\vec{r}_1)\right|^2|\alpha|^4 \\
&\quad -A_{\mathrm{LO}}(\vec{r}_1)A_{\mathrm{ph}}(\vec{r}_2)\overline{A_{\mathrm{LO}}(\vec{r}_2)}\,\overline{A_{\mathrm{ph}}(\vec{r}_1)}|\alpha|^4 \\
&\quad -4\overline{A_{\mathrm{LO}}(\vec{r}_1)}\,\overline{A_{\mathrm{ph}}(\vec{r}_2)}A_{\mathrm{LO}}(\vec{r}_2)A_{\mathrm{ph}}(\vec{r}_1)|\alpha|^4), \\
\langle\hat{E}_1^-(\vec{r}_1)\hat{E}_1^-(\vec{r}_1)\hat{E}_2^+(\vec{r}_2)\hat{E}_1^+(\vec{r}_1)\rangle &= 1/4\,|A_{\mathrm{LO}}(\vec{r}_1)|^2|A_{\mathrm{LO}}(\vec{r}_2)|^2|\alpha|^4 \\
&\quad +1/4\,|A_{\mathrm{LO}}(\vec{r}_1)A_{\mathrm{ph}}(\vec{r}_2)-A_{\mathrm{ph}}(\vec{r}_1)A_{\mathrm{LO}}(\vec{r}_2)|^2|\alpha|^2.
\end{aligned}
\tag{B8}
$$

Use of Eqs. B6-B8 enables the precise calculation of the required second order moment $\left\langle\widehat{\sigma}^2\right\rangle$ in terms of the modal shapes of photon and LO fields. Looking at the first row of Eq. B8, one observes that the term containing the highest power of $|\alpha|$ is: $\frac{1}{16}\left|A_{LO}(\vec{r}_1)\right|^4\left|A_{LO}(\vec{r}_2)\right|^4\left|\alpha\right|^8$, *identical* to the dominating term extracted from squaring Eq. B4, and meaning that it exactly cancels in the expression of the difference $\left\langle\widehat{\sigma}^2\right\rangle-\left\langle\widehat{\sigma}\right\rangle^2$, in which the leading remaining power is $\left|\alpha\right|^6$. The highest power for the variance $\Delta\widehat{\sigma}=\sqrt{\left\langle\widehat{\sigma}^2\right\rangle-\left\langle\widehat{\sigma}\right\rangle^2}$ will be therefore of the order of $\left|\alpha\right|^3$, and for the SNR the strongest dependence will be of the order of $\left|\alpha\right|$. An exact calculation of this SNR based on Eqs. B4, B6-B8 furnishes the graph of Fig. B1 and shows a fair linear dependence.

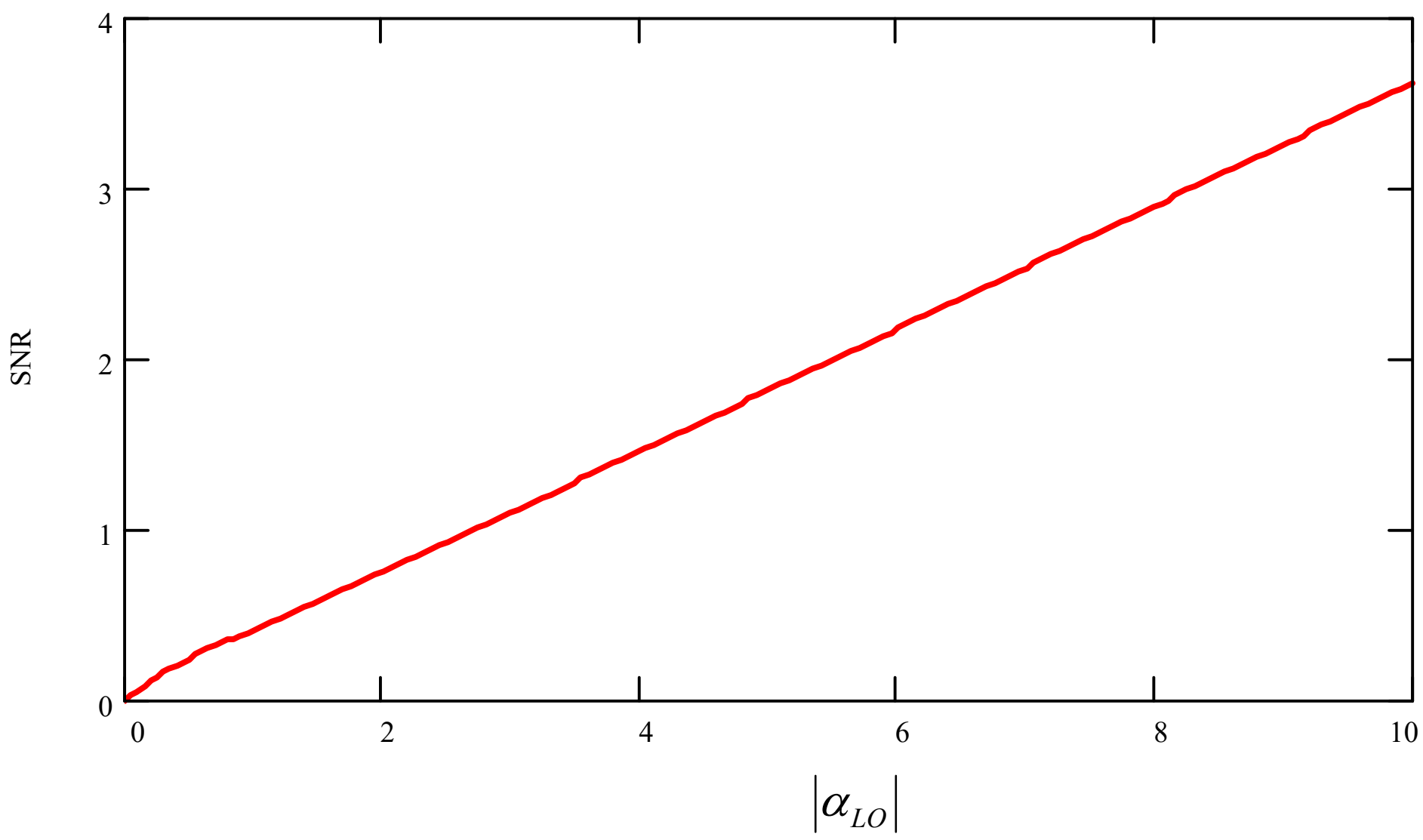

Fig B1. SNR as a function of field strength comparing the average *photon-LO* mixed signal to the MSD of the total signal.

## C. Interference visibility calculations for mixing between a single photon and a coherent state.

For open detectors encompassing most of the transverse modes, the fringe intensity visibility has been defined in both the time and space dimensions as the fractional reduction in correlation value at full overlap compared to its value at a non-overlapping (laterally displaced) situation [10,**Error! Reference source not found.**12]. This calculation for the case of mixing between a single photon and a coherent state is provided here, mainly with the purpose of validating the model applied in the article, associating the output of partial-aperture detectors with reported calculations for partial time overlap. Visibility for partial time overlap, as a function of the intensity was reported in Fig. 2 of [**Error! Reference source not found.**2] and was calculated there by a different method based on photon statistics considerations. Here it is simulated by the integration of Eq. (7) and depicted in Fig. C2 below, for varying lateral mode overlap. The figure also displays the depth of the Ou-Hong-Mandel (HOM) dip as a function of the intensity. Worth remarking is that although the visibility is reduced as a function of the LO strength, the absolute depth of the HOM dip increases, as evident from Fig. 2.

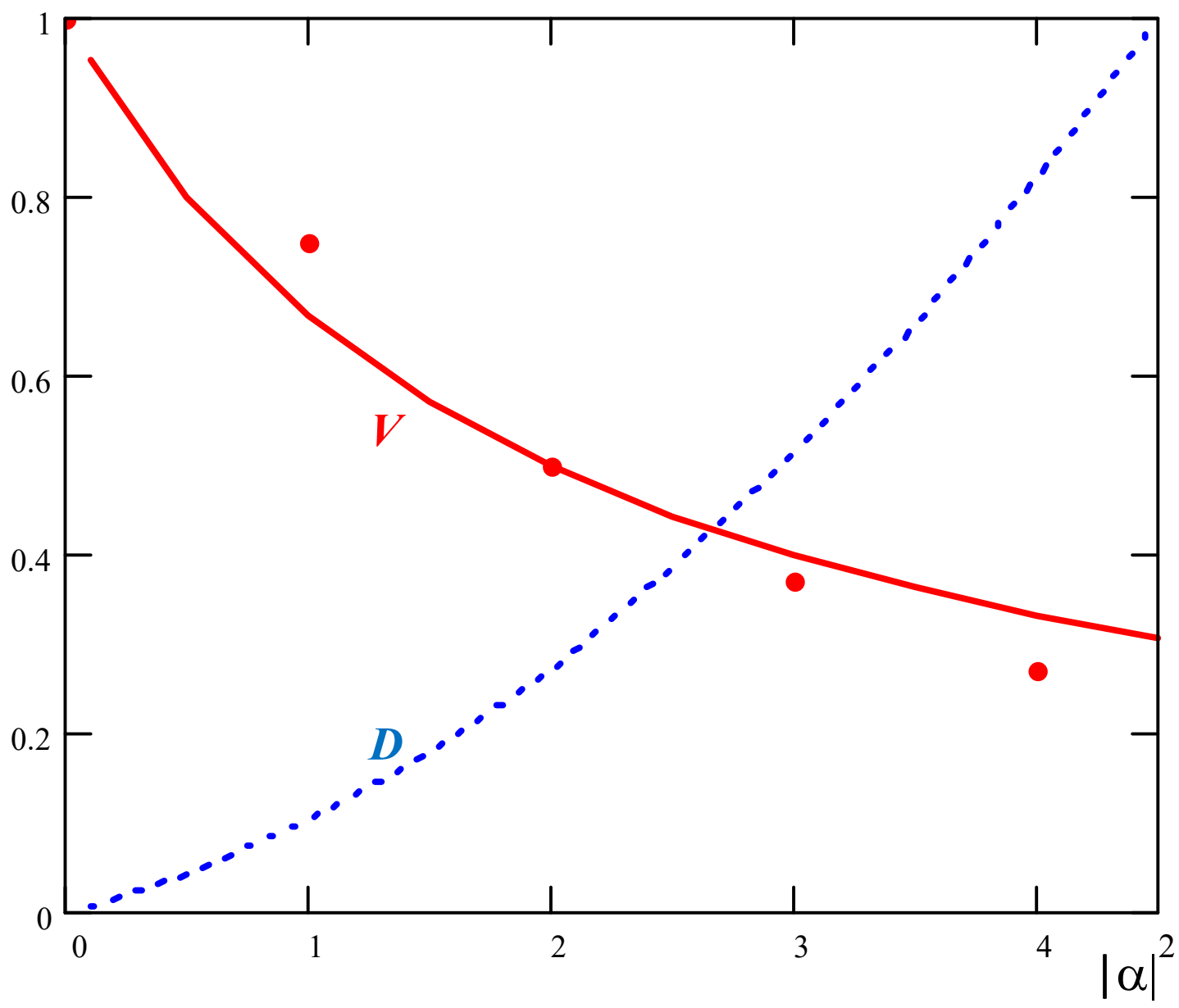

Fig. C2. Visibility (V) and depth (D) of the HOM dip for the mixing of a single photon with a coherent state as a function of the CS intensity parameter. The HOM depth was arbitrarily set to 1 for $|\alpha|$ = 4.5. The full points were selected from the interference visibility graph of Fig. 2 in ref. [7] in the article.

*ruschin@tauex.tau.ac.il